\documentstyle[prl,twocolumn,aps,epsf,rotate]{revtex}
\begin{document}

\title{On the dimensionality of spacetime}

\author{Max Tegmark}

\address{Institute for Advanced Study, Olden Lane,
Princeton, NJ 08540; max@ias.edu
}

\maketitle

\begin{abstract}              
Some superstring theories have more than one effective
low-energy limit, corresponding to 
classical spacetimes with different dimensionalities.
We argue that all but the 3+1-dimensional one might correspond to 
``dead worlds", devoid of observers, in which case
all such ensemble theories would
actually {\it predict} that we should find ourselves 
inhabiting a 3+1-dimensional spacetime.
With more or less than one time-dimension, 
the partial differential equations of nature would lack the hyperbolicity
property that enables observers to make predictions.
In a space with more than three dimensions, there can be no traditional 
atoms and perhaps no stable structures.
A space with less than three dimensions allows no 
gravitational force and may be too simple and barren
to contain observers.
\end{abstract}



\makeatletter
\global\@specialpagefalse
\def\@oddfoot{
\ifnum\c@page>1
  \reset@font\rm\hfill \thepage\hfill
\fi
\ifnum\c@page=1
  {\sl Published in Classical and Quantum Gravity, {\bf 14}, L69-L75 (1997).
  }\hfill
\fi
} \let\@evenfoot\@oddfoot
\makeatother


\section{INTRODUCTION}

\begin{figure}[phbt]
\centerline{{\vbox{\epsfxsize=8.8cm\epsfbox{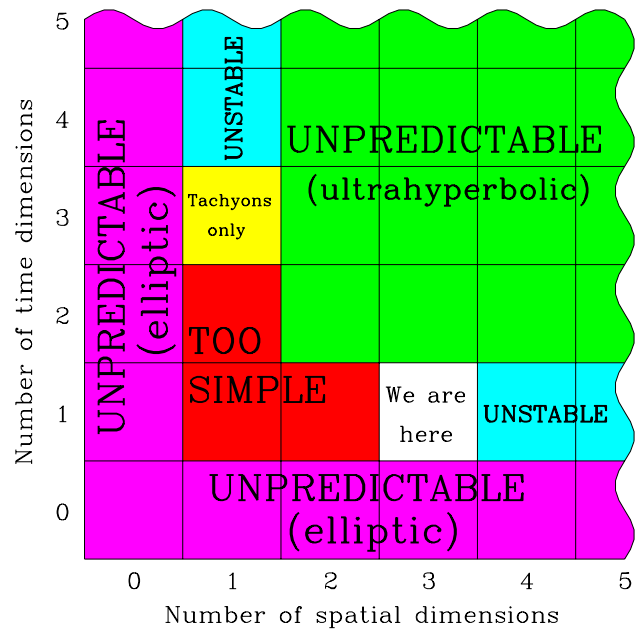}}}}
\smallskip
\caption{
\label{ExclusionFig}
When the partial differential equations of nature are elliptic or 
ultrahyperbolic, physics has no predictive power for an observer.
In the remaining (hyperbolic) 
cases, $n>3$ may fail on the stability requirement
(atoms are unstable) and $n<3$ may fail on the 
complexity requirement (no gravitational attraction, topological
problems).
}
\end{figure}

Many superstring theories have several stable 
(or extremely long-lived) states that constitute
different effective low-energy theories with different spacetime 
dimensionalities,
corresponding to different compactifications
of the many ({{\frenchspacing e.g.}}, 11 or 26) dimensions of the 
fundamental manifold.
Since the tunneling probabilities between these
states are negligible, such a theory for all practical 
purposes predicts an ensemble of classical 
$n+m$-dimensional spacetimes, and the
prediction for the dimensionality takes the form of
a probability distribution
over $n$ and $m$ \cite{Albrecht 1994}.
There are also inflationary models predicting a Universe consisting
of parts of exponentially large size having different
dimensionality \cite{Linde 1988}.
In this paper, we argue that 
this failure to make the unique prediction $(n,m)=(3,1)$ 
is {\it not} a weakness of such theories, but
a strength.
To compute the theoretically predicted probability
distribution for the dimensionality of our spacetime\footnote{
Here and thoughout, we let $n$ and $m$ refer to the number of
{\it non-compactified} space and time dimensions, or 
more generally to the effective spacetime
dimensionality that is 
relevant to the low-energy physics we will be discussing. 
},
we clearly need to take into account the selection
effect arising from the fact that some of
these states are more likely than others to
contain self-aware observers such as us. 
This is completely analogous to the familiar 
selection effect in cosmological galaxy surveys, where we must take
into account that bright galaxies are more likely 
than faint ones to be sampled \cite{Vilenkin}.
Below we will argue that if observers can only exist in a world
exhibiting a certain minimum complexity, predictability
and stability, then all such ensemble theories 
may actually predict that we should find ourselves 
inhabiting a 3+1-dimensional spacetime with 100\% certainty, 
as illustrated in Figure~\ref{ExclusionFig}, 
and that the Bayesean prior probabilities of quantum-mechanical 
origin are completely irrelevant.
We will first review some old but poorly known results regarding 
the number of spatial dimensions (when $m=1$), 
then present some new arguments regarding the number of time dimensions.
In both cases, we are {\it not} attempting to rigorously show that 
merely $(n,m)=(3,1)$ permits observers. Rather, we are simply arguing
that it is far from obvious that any other $(n,m)$ permits observers, 
since radical qualitative changes occur in all cases,
so that the burden of proof of the contrary
falls on the person wishing to criticize
ensemble theories with fine-tuning arguments.

\begin{figure}[bt]
\centerline{{\vbox{\epsfxsize=9.4cm\epsfbox{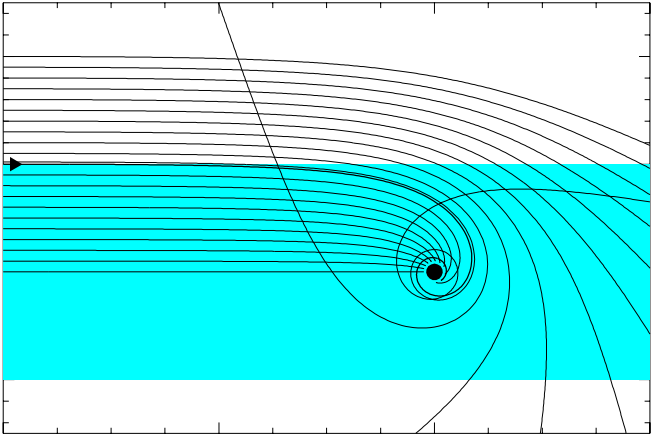}}}}
\smallskip
\caption{
\label{PlanetsFig}
The two body problem in four-dimensional space: 
the light particles that approach the heavy one at the center
either escape to infinity or get sucked into a cataclysmic collision.
There are no stable orbits.
}
\end{figure}

\section{WHY IS SPACE THREE-DIMENSIONAL?}

As was pointed out by Ehrenfest 
back in 1917 \cite{Ehrenfest 1917},
neither classical atoms nor planetary orbits can be stable
in a space with $n>3$, and traditional
quantum atoms cannot be stable either \cite{Tangherlini 1963}.
These properties are related to the fact that the 
fundamental Green function of the Poisson equation 
$\nabla^2\phi=\rho$, which gives the electrostatic/gravitational
potential of a point particle, is $r^{2-n}$ for $n>2$. 
Thus the inverse square law of electrostatics and gravity 
becomes an inverse cube law if $n=4$, {{\frenchspacing etc.}}
When $n>3$, the two-body problem no longer has any stable orbits
as solutions \cite{Buechel 1963}. 
This is illustrated in Figure~\ref{PlanetsFig},
where a swarm of light test particles are incident from
the left on a massive point particle (the black dot),
all with the same momentum vector but 
with a range of impact parameters.
There are two cases: those that start outside the shaded region
escape to infinity, whereas those with smaller impact parameters
spiral into a singular collision in a finite time.
We can think of this as there being a finite cross section for
annihilation. 
This is of course in stark contrast to 
the familiar case $n=3$, which gives either stable elliptic orbits or 
non-bound parabolic and hyperbolic orbits, and has no 
``annihilation solutions" except for the measure zero case
where the impact parameter is exactly zero.
A similar disaster occurs in quantum mechanics, where a study 
of the Schr\"odinger equation shows that the Hydrogen 
atom has no bound states for $n>3$ \cite{Tangherlini 1963}. 
Again, there is a finite
annihilation cross section, which is reflected by the fact 
that the Hydrogen atom has no ground state, but time-dependent 
states of arbitrarily negative energy.
The situation in General relativity is analogous
\cite{Tangherlini 1963}.
Modulo the important caveats in the discussion section,
this means that such a world cannot contain any objects that
are stable over time, and thus probably cannot contain
stable observers.

What about $n<3$? 
It has been argued \cite{Whitrow 1955} that organisms
would face insurmountable topological problems if $n=2$: 
for instance, two nerves cannot cross.
Another problem, emphasized by Wheeler \cite{MTW}, 
is the well-known fact 
(see {{\frenchspacing e.g.}} \cite{Deser 1984}) that there 
is no gravitational force in General Relativity with
$n<3$.
We will not spend more time listing problems with $n<3$, but simply 
conjecture that since 
$n=2$ (let alone $n=1$ and $n=0$) offers vastly less
complexity than $n=3$, worlds with $n<3$ are just too
simple and barren to contain observers.

\section{WHY IS TIME ONE-DIMENSIONAL?}

In this section, we will present an argument for why a  
world with the same laws of physics as ours and
with an $n+m$-dimensional spacetime can only 
contain observers if the number of time-dimensions $m=1$, 
regardless of the number of space-dimensions $n$.
Before describing this argument, which 
involves hyperbolicity properties of partial differential 
equations, let us make a few general comments about 
the dimensionality of time.

What would reality appear like to an observer in a manifold with 
more than one time-like dimension?
Even when $m>1$, there is no 
obvious reason for 
why an observer could not nonetheless {\it perceive} 
time as being one-dimensional, thereby 
maintaining the pattern of having ``thoughts" 
in a one-dimensional succession
that characterizes our own reality perception.
If the observer is a localized object, it will travel along 
an essentially 1-dimensional (time-like) world line
through the $n+m$-dimensional spacetime manifold. The  
standard General Relativity notion of its proper time is 
perfectly well-defined, and we would expect this to be the time that
it would measure if it had a clock and that it would subjectively
experience.

Needless to say, many aspects of the world would nonetheless
appear quite different. For instance, 
a re-derivation of relativistic mechanics for this more 
general case shows that energy now becomes an 
$m$-dimensional vector rather than a constant, whose
direction determines in which of the many time-directions
the world-line will continue, and in the non-relativistic
limit, this direction is a constant of motion. In other 
words, if two non-relativistic observers that are moving in 
different time directions happen to meet at a point in spacetime,
they will inevitably drift apart in separate
time-directions again, unable to stay together. 

Another interesting difference, which can be shown by an elegant 
geometrical argument \cite{Dorling 1969}, is that
particles become less stable when $m>1$. 
For a particle to be able to decay when $m=1$, 
it is not sufficient that there 
exists a set of particles with the same quantum numbers.
It is also necessary, as is well-known, that the sum of their rest 
masses should be less than the rest mass of the original 
particle, regardless of how great its kinetic energy may be.
When $m>1$, this constraint vanishes \cite{Dorling 1969}.
For instance, 
\begin{itemize}
\item 
a proton can decay into a neutron,  
a positron and a neutrino, 
\item
an electron can decay into a neutron, an antiproton and a neutrino,
and 
\item
a photon of sufficiently high energy can decay 
into any particle and its antiparticle.
\end{itemize}
In addition to these two differences, one can 
concoct seemingly strange occurrences involving 
``backward causation" when $m>1$.
Nonetheless, although such unfamiliar behavior may appear disturbing, 
it would seem unwarranted to assume that it would prevent any
form of observer from existing. After all, we must avoid the fallacy
of assuming that the design of our human bodies
is the only one that allows self-awareness. 
Electrons, protons and photons would still be stable if
their kinetic energies were low enough, so perhaps observers could
still exist in rather 
cold regions of a world with $m>1$\footnote{
It is, however, far from trivial to formulate a quantum field theory with a
stable vacuum state when $m>1$ \cite{Linde 1990}.}.
    
There is, however, an additional problem for observers 
when $m>1$, which has not been
previously emphasized even though the mathematical results
on which it is based are well-known.
If an observer is to be able to make any use of its self-awareness
and information-processing abilities, the laws of physics must be such that
it can make at least some predictions. Specifically, 
within the framework of a field theory, it should by 
measuring various nearby field values be able to
compute field values at some more distant space-time points 
(ones lying along its future world-line being particularly useful)
with non-infinite error bars.
If this type of well-posed causality were absent, then 
not only would there be no reason for observers to be self-aware,
but it would appear highly unlikely that 
information processing systems (such as computers and brains) 
could exist at all.

Although this predictability requirement may sound modest, 
it is in fact only met by a small class of partial differential 
equations (PDEs), essentially those which are hyperbolic.
We will now discuss the classification and causal structure of PDEs
in some detail.
This mathematical material is well-known,
and can be found in more detail in \cite{Courant & Hilbert}.
Given an arbitrary second order linear partial differential equation
in ${\bf R}^d$,
$$
\left[\sum_{i=1}^d \sum_{j=1}^d 
A_{ij} {\partial\over\partial x_i} {\partial\over\partial x_j}
+ \sum_{i=1}^d b_i {\partial\over\partial x_i}
+ c\right] u=0,
$$
where the matrix $A$ (which we without loss of generality 
can take to be symmetric), the vector $b$ and the scalar $c$
are given differentiable functions of the $d$ coordinates,
it is customary to classify it  
depending on the signs of the eigenvalues of $A$. The PDE is 
said to be
\begin{itemize}
\item
{\it elliptic} in some region of ${\bf R}^d$ if they are all positive 
or all negative there,
\item {\it hyperbolic} if one is positive and the rest are 
negative (or vice versa), and
\item {\it ultrahyperbolic} in the remaining case, {{\frenchspacing i.e.}},  
where at least two eigenvalues are positive and
at least two are negative.
\end{itemize}
What does this have to do with the dimensionality of spacetime?
For the various covariant field equations of nature that describe
our world (the wave equation $u_{;\mu\mu}=0$, 
the Klein-Gordon equation $u_{;\mu\mu} + m^2 u=0$, {\frenchspacing etc.}\footnote{
Our discussion will apply to matter fields with spin as well,
{\frenchspacing e.g.} fermions and photons,
since spin does not alter the causal structure of the solutions.
For instance, all four components of an electron-positron field 
obeying the Dirac equation satisfy the Klein-Gordon equation as well,
and all four components of the electromagnetic vector potential 
in Lorentz gauge satisfy the wave equation. 
}), the matrix $A$ will clearly have the same eigenvalues 
as the metric tensor.
For instance, they will be hyperbolic in a metric of 
signature $(+---)$, corresponding to $(n,m)=(3,1)$,
elliptic in a metric of signature $(+++++)$,
and ultrahyperbolic in
a metric of signature $(++--)$.

One of the merits of this standard 
classification of PDEs is that it determines their causal structure,
{\frenchspacing i.e.}, how the boundary conditions must be specified to make the
problem {\it well-posed}.
Roughly speaking, the problem is said to be well-posed if 
the boundary conditions determine a unique solution $u$ 
and if the dependence of this solution on the boundary
data (which will always be linear) is {\it bounded}.
The last requirement means that the solution $u$ at a given point 
will only change by a finite amount if the boundary data
is changed by a finite amount.
Therefore, even if an ill-posed problem can be formally 
solved, this solution would in practice be useless to an observer,
since it would need to measure the initial data with infinite
accuracy to be able to place finite error bars on the solution
(any measurement error would cause the error bars on the solution
to be infinite).

{\bf Elliptic} equations allow well-posed {\it boundary value problems}.
On the other hand, giving ``initial" data 
for an elliptic PDE on a non-closed hypersurface, say a plane, 
is an ill-posed problem. This means that
an observer in a world with no time dimensions (m=0) would
not be able do make any inferences at all about the situation
in other parts of space based on what it observes locally.
\begin{figure}[phbt]
\centerline{{\vbox{\epsfxsize=8.6cm\epsfbox{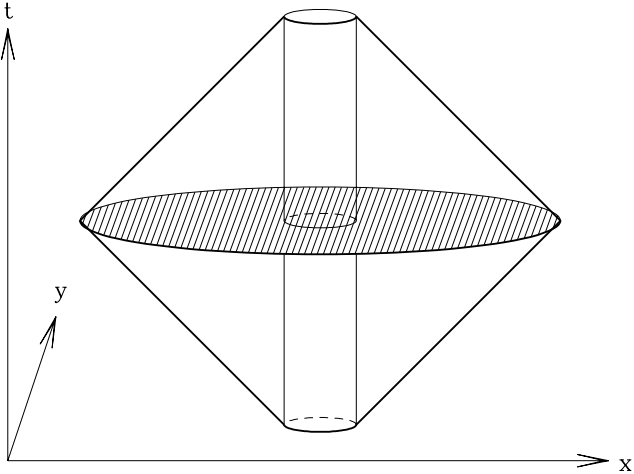}}}}
\smallskip
\caption{
\label{CausalityFig}
The causality structure for hyperbolic and ultra-hyperbolic equations.}

\end{figure}

{\bf Hyperbolic} equations, on the other hand, 
allow well-posed {\it initial-value problems}.
For example, specifying initial data ($u$ and $\dot u$)
for the Klein-Gordon equation
on the shaded disc in Figure~\ref{CausalityFig}
determines the solution in the volumes bounded by the
two cones, including the (missing) tips.
A localized observer can therefore make predictions about its
future. If the matter under consideration is 
of such low temperature that it is nonrelativistic,
then the fields will essentially contain only Fourier 
modes with wave numbers $|{\bf k}|\ll m$,
which means that for all practical purposes, the solution
at a point is determined by the initial data in a 
``causality cone" with an opening angle much narrower than
$45^\circ$.

In contrast, if the initial data for a hyperbolic PDE is 
specified on a hypersurface that is not spacelike, the problem 
becomes ill-posed. Figure~\ref{CausalityFig}, which is based on
\cite{Courant & Hilbert}, provides an intuitive 
understanding of what goes wrong.
A corollary of a remarkable theorem by Asgeirsson
\cite{Asgeirsson} is that if we specify $u$ in 
the cylinder in Figure~\ref{CausalityFig}, then this determines
$u$ throughout the region made up of the truncated double
cones. Letting the radius of this cylinder approach zero, 
we obtain the disturbing conclusion 
that providing data in a for all practical purposes
one-dimensional region determines the solution in a 
three-dimensional region. 
Such an apparent ``free lunch", where the solution seems to
contain more information than the input data,
is a classical symptom of ill-posedness. The price that must be
paid is specifying the input data with infinite accuracy,
which is of course impossible given real-world measurement 
errors. 
Clearly, generic boundary data allows no solution at all, since it is not 
self-consistent. 
It is easy to see that 
the same applies when specifying ``initial" 
data on part of a non-spacelike hypersurface,
{\frenchspacing e.g.}, that given by $y=0$.
These properties are analogous in $n+1$-dimensions, and illustrate
why an observer in an $n+1$-dimensional spacetime can
only make predictions in time-like directions.

Asgeirsson's theorem applies to the {\bf ultrahyperbolic} case as well,
showing that initial data on a hypersurface containing both spacelike and 
timelike directions leads to an ill-posed problem.
However, since a hypersurface by definition has 
a dimensionality which is one less than that of the 
spacetime manifold (data on a submanifold of lower dimensionality
can never give a well-posed problem), 
{\it there are no spacelike or timelike hypersurfaces} in the
ultrahyperbolic case, and hence no well-posed problems.
\footnote{The only remaining possibility is the rather contrived case
where data is specified on a null hypersurface.
To measure such data, an observer would need to ``live on the light cone", 
{{\frenchspacing i.e.}}, travel with the speed of light, which 
means that it would subjectively not
perceive any time at all (its proper time would stand still).
}
 
Since a mere minus sign distinguishes space from time,  
the remaining case $(n,m)=(1,3)$ is mathematically equivalent to 
the case where $(n,m)=(3,1)$ and all particles are 
tachyons \cite{Feinberg 1967} with imaginary rest mass.
Also in this case, an observer would be unable to make 
any predictions, since as described in more
detail in \cite{TOE},
well-posed problems require
data to be specified in the non-local region
{\it outside} the lightcones.

Above we discussed only linear PDEs, although the full system of coupled
PDEs of nature is of course non-linear. This in no way weakens our
conclusions about only $m=1$ giving well-posed initial 
value problems. When PDEs give ill-posed problems
even {\it locally}, in a small neighborhood of a hypersurface 
(where we can generically approximate 
the nonlinear PDEs with linear ones), it is obvious that 
no nonlinear terms can make them well-posed in a larger neighborhood.

\section{Discussion}

Our conclusions are graphically illustrated
in Figure~\ref{ExclusionFig}:
given the other laws of physics, it is not implausible that 
only a 3+1-dimensional spacetime can contain 
observers that are complex and stable enough to be able 
to understand and predict their world to any extent at all,
for the following reasons.
\begin{itemize}
\item More or less than 1 time dimension:
insufficient predictability.
\item More than 3 space dimensions:
insufficient stability.
\item Less than 3 space dimensions:
insufficient complexity.
\end{itemize}
Thus although application of the so-called weak anthropic principle 
\cite{Carter 1974} does in general {\it not} appear to 
give very strong predictions for 
physical constants \cite{Greenstein & Kropf}, 
its dimensionality predictions 
may indeed turn out to give the narrowest probability distribution 
possible. 
Viewed in this light, the multiple dimensionality prediction of some
superstring theories is a strength rather than a weakness, 
since it eliminates the otherwise embarrassing
discrete fine-tuning problem of having to explain
the ``lucky coincidence" that the compactification 
mechanism itself happened to single out only a 
3+1-dimensional spacetime. 

Needless to say, we have not attempted to rigorously demonstrate 
that observers are impossible for other dimensionalities. For instance,
within the context of specific models, 
one might consider exploring the possibility of stable structures
in the case $(n,m)=(4,1)$ based
on short distance quantum corrections to the $1/r^2$ potential or on
string-like (rather than point-like) particles. We have simply 
argued that it is far from obvious that any other combination 
than $(n,m)=(3,1)$ permits observers, since radical qualitative
changes occur when $n$ or $m$ are altered. For this reason,
a theory cannot be criticized for failing to predict a definitive spacetime 
dimensionality until the stability and predictability 
issues raised here have been carefully analyzed.

\acknowledgments
The author wishes to thank Andreas Albrecht, 
Dieter Maison, Harold Shapiro, John A. Wheeler,
Frank Wilczek and Edward Witten  
for stimulating discussions on some of the above-mentioned topics.



\end{document}